\documentclass[aps,pre,a4paper,twocolumn, showkeys]{revtex4}

\usepackage[]{amsmath}
\usepackage{amssymb}
\usepackage[dvips]{graphicx}
\usepackage{color}
\usepackage{bm}
\usepackage{tabularx}
\usepackage{mathtools}
\usepackage{algpseudocode}
\usepackage{enumitem}
\usepackage{multirow}
\usepackage{epstopdf}  


\newcommand{\rom}[1]{\uppercase\expandafter{\romannumeral #1\relax}}
\newcommand{\recheck}[1]{{#1}}

\newcommand{\vect}[1]{\textbf{\textit{#1}}}

\newcommand{\ins}{\textrm{ins}}

\begin{document}
\title{
DeePCG: constructing coarse-grained models via deep neural networks
}
\author{Linfeng Zhang}
\affiliation{Program in Applied and Computational Mathematics, 
Princeton University, Princeton, NJ 08544, USA}
\author{Jiequn Han}
\affiliation{Program in Applied and Computational Mathematics, 
Princeton University, Princeton, NJ 08544, USA}
\author{Han Wang}
\email{wang\_han@iapcm.ac.cn}
\affiliation{Institute of Applied Physics and Computational Mathematics,
Fenghao East Road 2, Beijing 100094, P.R.~China}
\affiliation{CAEP Software Center for High Performance Numerical
Simulation, Huayuan Road 6, Beijing 100088, P.R.~China}
\author{Roberto Car}
\affiliation{Department of Chemistry,  
Department of Physics, 
Program in Applied and Computational Mathematics, 
Princeton Institute for the Science and Technology of Materials,  
Princeton University, Princeton, NJ 08544, USA}
\author{Weinan E}
\email{weinan@math.princeton.edu}
\affiliation{Department of Mathematics and Program 
in Applied and Computational Mathematics, 
Princeton University, Princeton, NJ 08544, USA}
\affiliation{ Beijing Institute of Big Data Research, 
Beijing, 100871, P.R.~China}

\begin{abstract}
We introduce a general framework for constructing coarse-grained potential models without {\it ad hoc} approximations such as limiting the potential to two- and/or three-body contributions. 
The scheme, called Deep Coarse-Grained Potential (abbreviated DeePCG), exploits a carefully crafted neural network to construct a many-body coarse-grained potential. 
The network is trained with full atomistic data in a way that preserves the natural symmetries of the system. 
The resulting model is very accurate and can be used to sample the configurations of the coarse-grained variables in a much faster way than with the original atomistic model. 
As an  application we consider liquid water and use the oxygen coordinates as the coarse-grained variables, 
starting from a full atomistic simulation of this system at the {\it ab initio} molecular dynamics level. 
We find that the  two-body, three-body, and higher-order oxygen correlation functions produced by the coarse-grained and
full atomistic models
 agree very well with each other, illustrating the effectiveness of the DeePCG
 model on a rather challenging task.
\end{abstract}

\maketitle
\section{Introduction}
In molecular dynamics (MD), we are often faced with two types of 
coarse-graining tasks. 
In a first set of applications we are interested in evaluating the Landau free energy, which is a function of a small subset of coarse-grained (CG) variables. 
In this case the CG variables are either scalar or low dimensional vector variables. 
In a second set of applications we are interested in sampling with molecular dynamics (MD) or with Monte Carlo the configurations of an extensive set of CG variables. 
In this case the dimensionality of the CG space is proportional to the size of the system but is reduced relative to the full space of atomistic coordinates. 
The first type of CG variables is typically adopted to study problems like phase transitions,
where the objective is to perform detailed analyses of
the Landau free energy surface by finding the metastable states, the free energy
barriers between these states, the transition pathways, etc.
Take the melting of a solid as an example, the Steinhardt order parameters \cite{steinhardt1983bond} have been 
used as CG variables to differentiate solid (crystal) and liquid phases.
The second type of CG variables is typically used to accelerate configurational sampling relative to full atomistic simulations. 
For example, one may coarse-grain a polymer by replacing the monomers with point-like particles, or beads, connected by springs.

For a good description of the Landau free energy surface one needs to find good order parameters acting as CG variables and address the issues associated with crossing high energy barriers. 
\recheck{Typically these approaches are limited to a few CG variables, but recent work demonstrated that machine learning methods allow us to describe the functional dependence of the Landau free energy surface on several CG variables 
\cite{stecher2014free,mones2016exploration,lemke2017neural,galvelis2017neural,schneider2017stochastic,zhang2018reinforced,Zavadlav2018multiscale}. }
When considering extensive CG variables, the difficulty is often associated with finding an accurate free energy function in the space of the CG variables. 
Such free energy function usually depends on the CG variables in a complex and nonlinear way.
\recheck{
Therefore, finding a good representation of this function often requires substantial physical/chemical intuition
~\cite{lyubartsev1995calculation,rudd1998coarse, pagonabarraga2001dissipative, reith2003deriving, nielsen2004coarse, shinoda2008coarse, noid2008multiscale, noid2008multiscale2, shell2008relative,molinero2009water, 
larini2010multiscale,das2012multiscale,dinpajooh2017density,delyser2017extending,wagner2017extending,sanyal2016coarse,moore2016coarse}.}
In principle, machine learning methods can address this problem more accurately and in an automated way~\cite{behler2007generalized,bartok2010gaussian,schutt2017quantum,chmiela2017machine,han2017deep,zhang2018deep}, 
but most machine learning approaches so far have focused on the representation of the potential energy surface in the space of the atomistic degrees of freedom rather than the representation of the free energy surface in the space of the CG variables.
For example, the Deep Potential method \cite{han2017deep}, a model based on deep neural networks, has made it possible to parametrize an atomistic potential energy function derived from quantum mechanics without {\it ad hoc} approximations. 
A subsequent development of this approach, called Deep Potential Molecular Dynamics (DeePMD) \cite{zhang2018deep}, has allowed us to perform MD simulations of comparable quality to {\it ab initio} molecular dynamics (AIMD)~\cite{car1985unified} at the cost of classical empirical force fields. 

The free energy surface, rather than the potential energy surface, is the key physical quantity that we need to represent when dealing with CG variables.
In this work, we introduce the Deep Coarse-Grained Potential (DeePCG) scheme, an approach that generalizes the Deep Potential and DeePMD methods to representations of the free energy surface in the space of the CG variables, a quantity that will be called the CG potential in the following. 
\recheck{A related method to represent the many-body character of the CG potential in molecules was recently reported in Ref.~\cite{john2017many}.}
In our approach, similar to the Deep Potential and DeePMD methods, no {\it ad hoc} approximations are required, in addition to the network model itself, to represent the CG potential. 
The scheme is very accurate as demonstrated by the almost perfect agreement of the many-body correlations extracted from CG simulations with the corresponding correlations extracted from the original atomistic model. 
In the present work, we use liquid water as an example to illustrate the approach. 
We choose AIMD as the underlying atomistic model, and replace the individual water molecules with point-like particles located at the oxygen sites in the CG model. 
The excellent agreement of the second-, third-, and higher-order correlation functions between CG and atomistic models shows the promise of the DeePCG approach.

\section{Methodology}
\subsection{Basic Theory}
We consider a $d$-dimensional system with $N$ atoms in the constant-volume canonical ($NVT$) ensemble. 
The coordinates of the atoms,
in the laboratory frame, are $\bm{q}=\left\{q_1, q_2, \dots, q_{dN}\right\}\in\mathbb{R}^{dN}$.
The configurational distribution function is defined by
\begin{align}\label{eqn:dist-q}
  p(\bm q) = \frac{1}{Z} e^{-\beta V(\bm q)},  
\end{align}
where $Z= \int e^{-\beta V(\vect q)} \,d\bm q$ is the partition function.

The coarse-grained variables $\bm{\xi}(\bm{q})=\left\{\xi_1(\bm{q}), \xi_2(\bm{q}), \dots, \xi_{M}(\bm{q})\right\}$ are a reduced set of coordinates ($M<dN$).
$M$ can be finite and independent of the system size or it can be extensive with the system size, as in the two cases discussed in the introduction.
When $M$ is finite the CG variables are the so-called order parameters of the system.
When $M$ is extensive, the CG variables replace molecular objects with simpler sub-objects.
The configurational distribution of the CG system is the projection of the configurational distribution of the microscopic (atomistic) system onto the space of the CG variables:
\begin{align}\label{eqn:p-cg}
  p(\bm\xi) 
  = \frac 1{Z}  \int  e^{-\beta V(\vect q)} \delta (\bm\xi(\bm q) - \bm\xi) \,d\bm q.
\end{align}
The probability distribution in Eq.~\eqref{eqn:p-cg} allows us to define the CG potential and the forces acting on the CG variables as:
\begin{align}\label{eqn:u-def}
  U (\bm \xi) = - \frac 1\beta \ln p (\bm \xi),
\end{align}
and 
\begin{align}\label{eqn:cg-f-def}
  \bm F (\bm \xi) = -\nabla_{\bm\xi} U(\bm\xi),
\end{align}
respectively.
Eq.~\eqref{eqn:u-def} tells us that a good CG potential should reproduce accurately the \emph{full} configurational distribution of the CG variables in the atomistic model.
Testing the quality of the full configurational distribution of the CG variables is difficult, and, typically tests have been based only on two- and three-body correlation functions~\cite{noid2008multiscale,noid2008multiscale2,larini2010multiscale,das2012multiscale}.  

$U(\bm{\xi})$ is uniquely specified by the underlying atomistic model and the definition of the CG degrees of freedoms. 
Therefore, constructing a CG model involves two steps:
(1) the choice of an appropriate CG-potential representation, and 
(2) the optimization of the parameters that define the potential representation.
The way in which these two issues are addressed differentiates alternative schemes.

We notice that, even if we knew exactly $U(\bm{\xi})$, we would not have a closed deterministic form for the equation of motion of the CG variables due to the $dN-M$ missing degrees of freedom in the CG potential.
The issue of the dynamics of the CG variables has been addressed in the literature.
See, e.g., Refs.~\cite{hijon2010mori,lu2014exact}.
Further assumptions, like a time-scale separation between the CG variables and the remaining degrees of freedom, are usually required to recover dynamical information of the atomistic model. 
In the following we shall focus on the accurate construction of the CG potential. We will leave to future studies the investigation of CG dynamics. 

\subsection{CG potential representation}
We adopt a neural network representation $U^{\bm{w}}(\bm{\xi})$ for the CG potential $U(\bm{\xi})$. 
Here $\bm{w}$ are the parameters to be optimized by the training process.
$U^{\bm{w}}(\bm{\xi})$ should be constructed using in input only the generalized coordinates $\bm{\xi}$, without any human intervention in the optimization process.
The $U^{\bm{w}}(\bm{\xi})$ constructed in this way should preserve the symmetry properties of $U(\bm{\xi})$.
In this manuscript we limit ourselves to considering CG objects
that behave as point particles and have only positional dependence.
In this case, the $\bm \xi$ variables are the coordinates of the CG particles.
More general choices of the CG objects have been suggested in the literature~\cite{voth2008coarse,underhill2004coarse,goodchild2007structural,bhargava2009formation} when dealing with, e.g., polymers, biological molecules, or colloidal particles. 
In these cases it may be useful to consider rods, ellipsoids, particles connected by springs, etc., as the CG objects. In principle, all these cases 
could be treated with the present formalism. In the setup adopted here of point-like CG objects, the CG potential $U^{\bm{w}}$ is extensive, intrinsically many-body, and should preserve the translational and rotational invariance, as well as the permutational symmetry of the CG objects.

\begin{figure}[tbp]
\centering
 \includegraphics[width=0.48\textwidth]{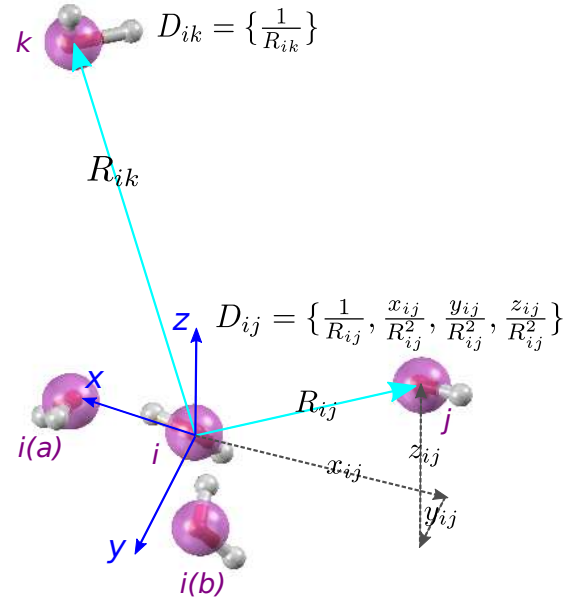}
 \caption{Schematic plot of the neural network input
  for the environment of CG particle $i$, using water as an example.
  Red and white balls represent the oxygen and the hydrogen atoms of the microscopic system, respectively.
  Purple balls denote CG particles, which, in our example, are centered at the positions of the oxygens.
  We first sort all the CG particles within the cutoff radius
  $R_c$ centered at $i$, 
  according to their inverse distances from $i$. 
  These particles constitute the neighbors of $i$.
  $i(a)$ and $i(b)$ are the first and the second nearest neighbor, respectively,
  of $i$.
  $j$ and $k$ are generic neighbors of $i$.
  $i$, $i(a)$, and $i(b)$ define the local frame of $i$.
  In this local frame, $i$ is the origin;
  the arrow from $i$ to $i(a)$ defines the $x$ axis;
  the directional normal to the plane containing $i$, $i(a)$, and $i(b)$ defines the $z$ axis;
  the $y$ axis is then assigned with the right-hand rule.
  Site $j$ is close to $i$ and is described with full radial and angular information by the descriptor $\vect D_{ij} = \{1/ R_{ij}, x_{ij}/R^2_{ij}, 
    y_{ij}/R^2_{ij}, z_{ij}/R^2_{ij}\}$, 
    where $(x_{ij}, y_{ij}, z_{ij})$ are Cartesian coordinates in the local frame of $i$. 
    Site $k$ is far from $i$ and is described with radial information only by the descriptor $\vect D_{ik} = \{1/ R_{ik}\}$.}
 \label{fig:plot_all}
\end{figure}

\begin{figure}[ht!]
\centering
 \includegraphics[width=0.48\textwidth]{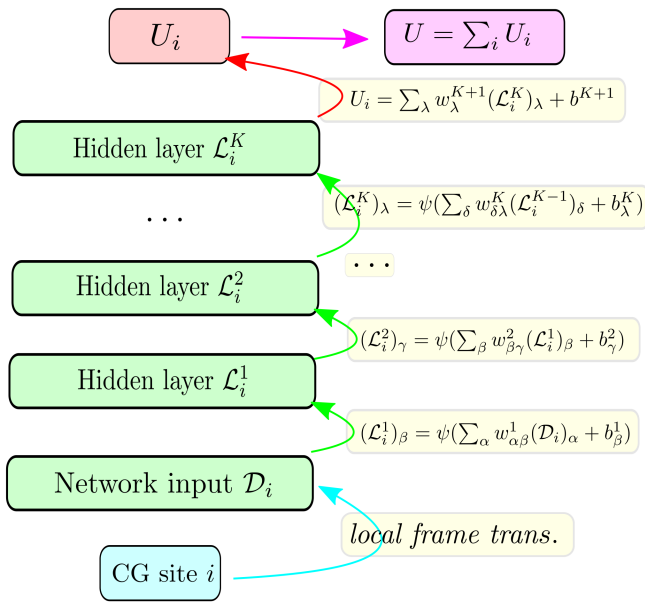}
 \caption{Schematic plot of the sub-network structure for the CG particle $i$.
  $\mathcal{D}_i$ (see definition in text) is the input and $U_i$ is the output.
In this sub-network, data flow from the input layer ($\mathcal{D}_i$) 
to the output layer ($U_i$) through $K$ hidden layers,
where each layer is a composition of a linear transformation and a piecewise nonlinear operation $\psi(\cdots)$.
We use the hyperbolic tangent for the nonlinear function $\psi$.
This procedure is adopted for all the hidden layers. 
In the final step going from the last hidden layer to $U_{i}$, 
only the linear transformation is applied.}
 \label{fig:nn_struc}
\end{figure}

All the properties of the CG potential described above are preserved by the Deep Potential model \cite{han2017deep}. 
To illustrate how it works, we use the example of liquid water. 
We write the CG potential as a sum of the local contributions of the CG particles, i.e., $U^{\bm w}(\bm\xi)=\sum_i U_{i}^{\bm w}(\bm\xi)$. $U_{i}^{\bm w}(\xi)$, the potential contribution of the CG particle $i$, is constructed in two steps.
First, the coordinates of the CG particle $i$ and its neighbors within a cut-off radius $R_c$ are transformed
into the descriptors $\{\vect D_{ij}\}$ of the local environment of the CG particle $i$.
We call this procedure local frame transformation and refer to Fig.~\ref{fig:plot_all} for more details. 
In the following we use the symbol $\mathcal{D}_i$ to denote the entire set of descriptors for atom $i$.
Next, as illustrated in Fig.~\ref{fig:nn_struc}, the descriptors $\mathcal{D}_i$ are given in input to 
a fully connected feedforward neural network to compute the potential contribution of the CG particle $i$.
The mathematical formulation of the network structure is also presented in Fig.~\ref{fig:nn_struc},
where the operation of each layer of the network corresponds to a linear mapping of the output
from the previous layer combined with a nonlinear mapping. 
The translational and rotational symmetries are preserved by the local frame transformation.
The permutational symmetry is preserved because:
(a) for each CG particle $i$, its descriptors $\vect D_{ij}$ are sorted in ascending order according to the inverse distances between particles $i$ and $j$; 
(b) the subnetworks associated with the same type of particles share the same parameters $\bm w$; 
(c) $U^{\bm w}(\bm\xi)=\sum_i U_{i}^{\bm w}(\bm\xi)$ is an additive relationship. 
More details on the Deep Potential method can be found in Refs.~\cite{han2017deep,zhang2018deep}. 
Due to the adoption of a finite cutoff radius, the simulation cost of the DeePCG model scales linearly with the system size.

\subsection{CG potential optimization}
The construction of the CG potential $U^{\bm w}(\bm\xi)$, introduced in Subsection B, has many similarities with the construction of the potential energy $V(\bm q)$, using the DeePMD method.
There is, however, a very important difference in the two cases. 
In the DeePMD case the potential energy $V(\bm q)$ is directly available from the underlying AIMD simulations.
In the DeePCG case, the CG potential is a free energy and is not directly available.
Therefore, the optimization for the CG potential requires a specific formulation. 
\recheck{
We adopt a force-matching scheme like the one in the multi-scale coarse-graining method introduced by Voth et al.~\cite{noid2008multiscale}.
In addition, we pay special attention to the fact that a neural network representation $U^{\bm w}(\bm\xi)$ may have tens of thousands, or more, variational parameters, and a suitable optimization algorithm is needed.}

A straightforward force-matching approach would consist in fitting accurate mean forces from atomistic simulations.
There have been many efforts in this direction  \cite{ciccotti2005blue,maragliano2006temperature,abrams2008efficient}.
Of particular interest is a simple formula proposed by Ciccotti et al.~\cite{ciccotti2005blue},
in which a set of $dN$-dimensional vectors $\bm{b}_j(\bm{q})$ that satisfy
\begin{align}\label{eqn:b-cond}
\nabla_{\bm{q}}\xi_i(\bm{q})\cdot\bm{b}_j(\bm{q})=\delta_{ij}, \quad i,j = 1,\dots M,
\end{align}
is introduced.
Then the mean force on $\xi_i(\bm{q})$, namely the negative gradient of $U(\bm{\xi})$ with respect to the position of the $i$-th CG particle, can be expressed as
\begin{align}\label{eqn:cg-f-comp}
  F_{i}(\bm \xi)
  =
  -\partial_i U(\bm \xi)
  =
  \langle \mathcal{F}_{i}(\bm{q}) \rangle_{\bm{\xi}=\bm{\xi}(\bm{q})}, 
\end{align}
with an instantaneous force estimator
\begin{align}\label{eqn:cg-f-esti}
  \mathcal{F}_{i}(\bm{q})= - \bm{b}_i(\bm{q})\cdot\nabla_{\bm{q}}V(\bm{q}) + \frac{1}{\beta}\nabla_{\bm{q}}\cdot \bm{b}_i(\bm{q}).
\end{align}
Here $\langle \cdots \rangle_{\bm{\xi}=\bm{\xi}(\bm{q})}$ denotes conditional expectation over the
equilibrium distribution of the system restricted to the hypersurface $\bm{\xi}=\bm{\xi}(\bm{q})$.
To train the DeePCG model one needs to minimize the so-called loss function with respect to the model parameters $\bm w$.
The most natural choice of loss function in terms of force-matching is
\begin{align}\label{eqn:loss}
  \hat{L}(\bm w) =
  \frac 1{dDM} \sum_{n=1}^{D} \sum_{i=1}^{dM}
  \big\vert F_i(\bm\xi_n) + \partial_i U^{\bm w}(\bm \xi_n) \big\vert^2,
\end{align}
where $D$ is the number of configurations of CG variables $\bm\xi_n$ in the dataset
and the mean force $F_i(\bm\xi_n)$ is estimated with Eq.~\eqref{eqn:cg-f-comp}.
\recheck{
We notice that the sample of CG configurations in Eq.~\eqref{eqn:loss} is very general in the sense that it does not need to be an equilibrium sample at the thermodynamic conditions of the atomistic simulation. 
For example, the sample could include, in an enhanced way, accessible CG configurations that have a small probability of occurrence at the thermodynamic conditions of interest, such as in the case of rare events. 
We stress, however, that the sampling of the microscopic degree of freedom orthogonal to $\bm\xi$ in Eq.~\eqref{eqn:cg-f-comp} must be done at the appropriate thermodynamic conditions. 
In practice, the different configurations $\bm\xi_n$ in the dataset can be extracted from unconstrained MD or Monte Carlo (MC) simulations of the microscopic atomistic model at different temperatures.}

The above straightforward approach is not convenient when the conditional expectation values in Eq.~\eqref{eqn:cg-f-comp} require computationally expensive constrained/restrained simulations.
In this situation we find it more convenient to approximate the ensemble average $\langle\cdots\rangle_{\bm q}$ with the average ($\frac{1}{D}\sum_{n=1}^{D}\cdots$) over the configurations $\bm \xi_n$ (see Eq.~\eqref{eqn:loss-ins1} below). 
\recheck{
The latter average does not require constrained/restrained simulations,
but it requires $\bm \xi_n$ to be extracted from equilibrium atomistic simulations at the temperature selected in Eq.~\eqref{eqn:dist-q}.
Then the mean force $F_i$ in the loss function~\eqref{eqn:loss} can be replaced by the instantaneous force $\mathcal F_i$. }
In other words, this corresponds to using an instantaneous version of the loss function
\begin{align}\label{eqn:loss-ins}
   \hat{L}^\ins(\bm w) =
  \frac 1{dDM} \sum_{n=1}^{D} \sum_{i=1}^{dM}
  \big\vert \mathcal F_i(\bm\xi_n) + \partial_i U^{\bm w}(\bm \xi_n) \big\vert^2.
\end{align}
With a sufficiently large representative dataset, we expect that the ensemble average
of the difference between predicted and instantaneous forces should be approximated quite well by $\hat{L}^\ins(\bm w)$, i.e.:
\begin{align}\label{eqn:loss-ins1}
  L^\ins(\bm w)
  &\coloneqq
  \frac 1{dM} \sum_{i=1}^{dM}
  \big\langle
  \big\vert\mathcal F_i(\bm \xi(\bm q)) + \partial_i U^{\bm w}(\bm \xi(\bm q)) \big\vert^2
  \big\rangle_{\bm q}\\
  &\approx\hat{L}^\ins(\bm w) \notag.
\end{align}
This amounts to an ergodicity requirement for the atomistic system and is always valid if the system samples an equilibrium thermodynamic state.

By definition, $\hat{L}^\ins(\bm w)$ is much easier to compute than $\hat{L}(\bm w)$. Below we argue that $\hat{L}^\ins(\bm w)$ is also a valid loss function to optimize CG potential. 
To see this, note that the instantaneous force can be viewed as the mean force  plus a random error $R$, which depends on the microscopic configuration $\bm q$, i.e.,
\begin{align}\label{eqn:insf-error}
  \mathcal F_i (\bm q) = F_i (\bm \xi(\bm q)) + R_i(\bm q).
\end{align}
By using Eq.~\eqref{eqn:cg-f-comp}, the average $\langle R_i(\bm q)\rangle_{\bm\xi=\bm\xi(\bm q)}$ in the constrained ensemble vanishes,
so the average $\langle R_i(\bm q)\rangle_{\bm q}$ also vanishes.
By inserting \eqref{eqn:insf-error} into \eqref{eqn:loss-ins1}, the instantaneous loss function \eqref{eqn:loss-ins1} becomes 
\begin{align}\label{eqn:loss-ins2}
  L^\ins(\bm w) = L(\bm w) 
  +
  \frac 1{dM} \sum_{i=1}^{dM}
  \big\langle
  R^2_i(\bm q)  
  \big\rangle_{\bm q},
\end{align}
with
\begin{align}
  L(\bm w)\coloneqq
  \frac 1{dM} \sum_{i=1}^{dM}
    \big\langle
    \big\vert F_i(\bm \xi(\bm q)) + \partial_i U^{\bm w}(\bm \xi(\bm q))\big\vert^2
    \big\rangle_{\bm q}.
\end{align}
Since the second term on the right hand side of Eq.~\eqref{eqn:loss-ins2} is independent of $\bm w$, 
$L^\ins(\bm w)$ and $L(\bm w)$ have the same minimizer. 
This equivalence justifies our usage of $\hat{L}^\ins(\bm w)$ as the loss function.

\recheck{
In the application example that we discuss in the next section we use CG variables that depend linearly on the microscopic coordinates, similar to  Ref.~\cite{noid2008multiscale}.
However, the method that we have illustrated in Eqs.~\eqref{eqn:cg-f-comp} and~\eqref{eqn:cg-f-esti} can deal with non-linear dependencies as well, as in the method discussed in Ref.~\cite{kalligiannaki2015geometry}.
}

In practice, we find that the stochastic gradient descent method is very efficient to optimize loss function \eqref{eqn:loss-ins},
which is a highly non-convex function corresponding to a rugged landscape in the large parameter space due to the nonlinearity of the neural network interpolation.
This ruggedness does not seem to constitute an essential difficulty since the different local minima found with the stochastic gradient descent (SGD) method approximate equally well the physics associated to the target function.
We will discuss this issue in more detail later.
Within our approach, the stochastic gradients $\nabla_{\bm w}l(\bm w)$, applied to update the parameters at each step, are provided by the average over a small batch $\mathcal B$, a subset of the whole dataset:
\begin{align}\label{eqn:loss-batch}
  l(\bm w)\coloneqq\frac 1{dM} \sum_{i=1}^{dM}
  \frac1{\vert\mathcal B\vert}\sum_{\alpha \in \mathcal B}
  \big\vert \mathcal F_i(\bm \xi(\bm q_\alpha)) + \partial_i U^{\bm w}(\bm \xi(\bm q_\alpha)) \big\vert^2, 
\end{align}
where $\vert\mathcal B\vert$ denotes the batch size. 
The above procedure is different from the scheme adopted in Ref.~\cite{noid2008multiscale}, 
in which the full gradients $\nabla_{\bm w}{\hat{L}^\ins(\bm w)}$ are applied to update the parameters at each step. 
We find that SGD greatly reduces the number of gradient evaluations that are required.

\recheck{We note that other systematic procedures to optimize the parameters have been discussed in the literature. 
Of particular interest is the iterative Boltzmann inversion method \cite{reith2003deriving},
which works by iteratively optimizing the CG interactions until the radial distribution functions of the CG system match those of the target atomistic simulation.
By construction, it provides accurate two-body correlations.}

\section{Coarse-graining of liquid water}
To show how we construct the CG potential for an extensive CG system,
we use the coarse graining of a liquid water model from an {\it ab initio} density functional theory (DFT)~\cite{kohn1965self} based simulation into effective ``water particles'' as an example.
Because of its importance as a solvent in chemical and biological systems and its unique properties, the study of water is of wide interest.
The DFT potential energy surface is intrinsically many-body.
Developing an accurate CG model that represents a water molecule by a single particle is an ever-evolving and ongoing quest
\cite{noid2008multiscale,noid2008multiscale2,larini2010multiscale,das2012multiscale,lyubartsev1995calculation,reith2003deriving,shell2008relative,molinero2009water}. 

Constructing effective interactions to achieve this goal has usually required a large amount of human effort combined with substantial physical/chemical intuition.
For example, in the mW monatomic potential~\cite{molinero2009water}, which has been successfully used to study crystallization of water \cite{moore2011structural},
a specially designed Lennard-Jones-like form is used for two-body interactions while three-body interactions are adapted from the Stillinger-Weber potential \cite{stillinger1985SW}.
In principle, coarse graining approaches that do not require physical/chemical intuition are possible by exploiting general variational principles, such as the one adopted in the multi-scale coarse-graining (MS-CG)~\cite{noid2008multiscale} or the iterative Boltzmann inversion~\cite{reith2003deriving} methods. 
\recheck{
However, even when using general variational principles to bridge atomistic and CG scales, the faithfulness of distribution of the CG variables may still depend on the CG representation. 
For example, in the applications of the MS-CG method to liquid water, the two- and three-body distribution functions of the CG variables still show non-negligible deviations from the corresponding target microscopic distributions~\cite{larini2010multiscale,das2012multiscale}.}

The DFT dataset in our example comes from Ref.~\cite{distasio2014individual}. The electronic structure of the water system is modeled by DFT with the PBE0 exchange-correlation functional \cite{Carlo1999PBE0}
and includes the long-range dispersion interactions self-consistently using the Tkatchenko-Scheffler model \cite{TS2009TS}.
The corresponding AIMD simulation \cite{car1985unified} adopts periodic boundary conditions 
and deuterons replace protons for a larger integration time step (0.5~fs).
The simulation data consist of snapshots from a 20~ps-long trajectory in the NVT
ensemble, where $N=192$ (64 H${}_2$O molecules), $V=1.9275~\text{nm}^3$ (simple cubic periodic simulation cell), and $T=330$~K.
In total 40,000 snapshots are recorded.
\recheck{
The important difference between training DeePMD and DeePCG is that in DeePCG one is attempting to estimate mean forces that correspond to conditioned averages of fluctuating atomic forces for fixed CG configurations, while in DeePMD one is attempting to estimate the deterministic atomic force in a fixed atomic configuration. 
Therefore, a short AIMD trajectory is not sufficient to train a DeePCG model with satisfactory accuracy,
but this difficulty is circumvented by constructing a DeePMD model \cite{zhang2018deep} from the AIMD data and sampling the configurations with a much longer DeePMD trajectory (15~ns).}
Figs.~\ref{fig:rdf}, \ref{fig:adf}, and \ref{fig:qs} compare DeePMD and AIMD configurations 
in terms of the O-O radial distribution function (RDF), O-O-O angular distribution functions (ADFs), and the distributions of two averaged local Steinhardt parameters (defined in Appendix~\ref{app:a})~\cite{lechner2008accurate}, respectively.
It is observed that the configurations sampled by DeePMD are in almost perfect agreement with the AIMD data.
Therefore, when considering the oxygen configurations, training with the data generated by DeePMD is essentially indistinguishable from that with data generated by AIMD.

Now we construct the DeePCG model. We use oxygen as the CG particle.
We define the local environment of an O atom with the same cutoff radius adopted in the DeePMD model, i.e., $R_c=6$\AA.
We use the full radial and angular information
for the 16 CG particles closest to the particle at the origin 
(see, e.g., particle $j$ in Fig. \ref{fig:plot_all}), 
while retaining only radial information for all the other particles 
within $R_c$,
(see, e.g., particle $k$ in Fig. \ref{fig:plot_all}).
Next, the local environment of each CG particle defines a sub-network, and we use 4 hidden layers with decreasing number of nodes per layer,
i.e., 120, 60, 30, and 15 nodes from the innermost to the outermost layer,
to construct the corresponding contribution to the CG potential.

The training process minimizes $\hat{L}^\ins(\bm w)$ defined in Eq.~\eqref{eqn:loss-ins}.
The force on each oxygen in the atomistic model serves as the instantaneous estimator $\mathcal{F}_{i}$ in Eq.~\eqref{eqn:cg-f-esti}.
We employ the stochastic gradient descent method with the Adam optimizer~\cite{kingma2014adam}
to update the parameters of each layer,
with a batch size of 4 and a learning rate that exponentially decays with the training step.
In our current implementation, the training process requires
15 hours on a ThinkPad p50 laptop computer with
an Intel Core i7-6700HQ CPU and 32 GB memory.
The DeePMD-kit~\cite{wang2017deepmd} is used for optimizations and MD simulations of both the DeePMD and the DeePCG models.

\recheck{
After training, we perform an NVT simulation on the CG variables. The initial snapshot for this simulation is taken directly from a snapshot selected along the AIMD trajectory.
The CG force is generated directly by analytical gradient of the CG potential, the volume and the temperature are the same of the AIMD simulation, and the temperature is controlled using a Langevin thermostat with a damping time $\tau= 0.1$ps.
In addition, using the same strategy, we perform an NVT simulation on 512 CG variables,
where the only difference is that the number of CG variables and the size of the simulation region are 8 times larger than those of the AIMD simulation.
}

\section{Discussion}
\begin{figure}[]
\centering
 \includegraphics[width=0.4\textwidth]{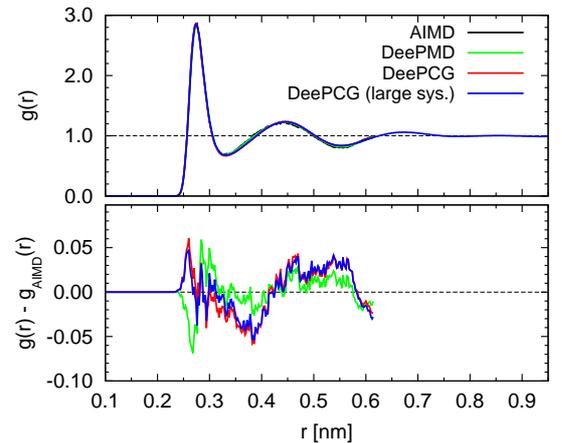}
 \caption{Upper panel: the O-O RDFs of liquid water from AIMD and DeePMD for a system with 64 water molecules, and from DeePCG simulation for systems with 64 and 512 CG particles; lower panel: the deviations of DeePMD and of two DeePCG models relative to the AIMD result.}
 \label{fig:rdf}
\end{figure}

\begin{figure}[]
\centering
    \includegraphics[width=0.4\textwidth]{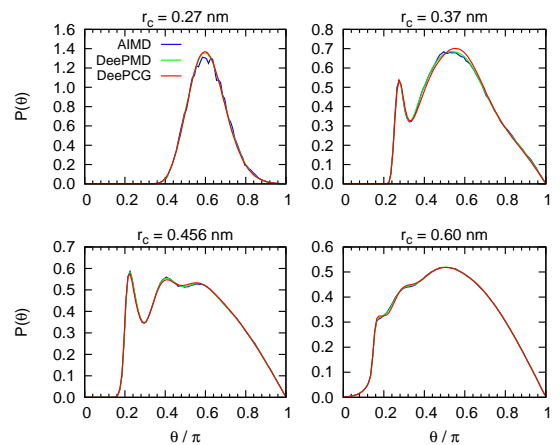}
 \caption{The O-O-O ADFs of liquid water from AIMD, DeePMD, and DeePCG simulations. The results for four different cutoff radii are provided.}
 \label{fig:adf}
\end{figure}

\begin{figure}[]
\centering
  \includegraphics[width=0.4\textwidth]{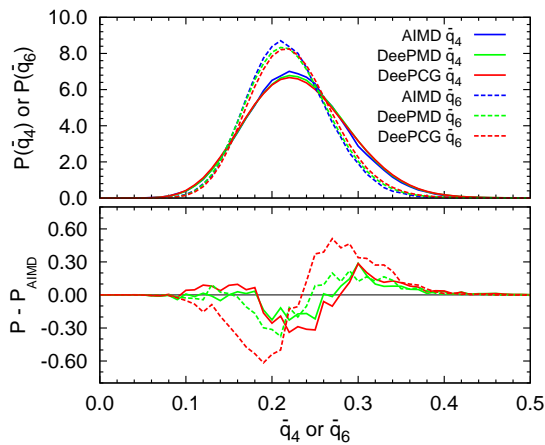}
 \caption{Upper panel: the $\bar q_4$ and $\bar q_6$ distribution function of liquid water from AIMD, DeePMD, and DeePCG simulations; lower panel: deviations of DeePMD and of DeepCG from the AIMD results.}
 \label{fig:qs}
\end{figure}

In Figs.~\ref{fig:rdf}, \ref{fig:adf}, and \ref{fig:qs}, we show that the DeePCG model reproduces very well the 
oxygen correlation functions of the atomistic DeePMD model and, by extension, of the underlying AIMD model.
In addition to comparing 2- and 3-body correlations, as done in standard protocols~\cite{larini2010multiscale,das2012multiscale}, 
we also perform tests on how well the DeePCG model preserves higher order distribution properties. 
In this regard, we calculate the sample averaged local Steinhardt bond order parameters
$\bar q_4$ and $\bar q_6$, and find satisfactory agreement between the DeePCG and DeePMD models.


In the example that we discussed above we use $\hat{L}^{ins}(\bm w)$ to optimize a CG model of water.
We find that to base the optimization on $\hat{L}(\bm w)$ defined in Eq.~\eqref{eqn:loss} is significantly less efficient. 
This is because when the oxygens are the CG variables, very long constrained simulations using Eq.~\eqref{eqn:cg-f-comp} are required to sample exhaustively the allowed configurations of the hydrogen bond network (HBN).
Typically, when the oxygen positions are fixed, as in a constrained simulation, different HBN configurations are compatible with the fixed oxygen configurations, 
but it takes a very long time, typically of the order of a few nanoseconds, for the system to sample different HBN configurations.
This is because of the long-range correlations imposed on the HBN by the Pauling ice rules (i.e., each oxygen has two nearer and two more distant hydrogen neighbors)~\cite{pauling1928shared}.
Thus, the scheme used here for matching the on-the-fly instantaneous forces is much more efficient.

It is well-known that neural network models are highly nonlinear functions of the parameters $\bm w$. 
Multiple local minima exist in the landscape of the loss functions $L(\bm w)$ or $L^{\ins}(\bm w)$.
Indeed, different initializations of the weights often lead to different local minimizers of the loss function.
This, however, does not seem to be a serious problem as demonstrated by the test described below. 

\begin{figure}[]
\centering
  \includegraphics[width=0.4\textwidth]{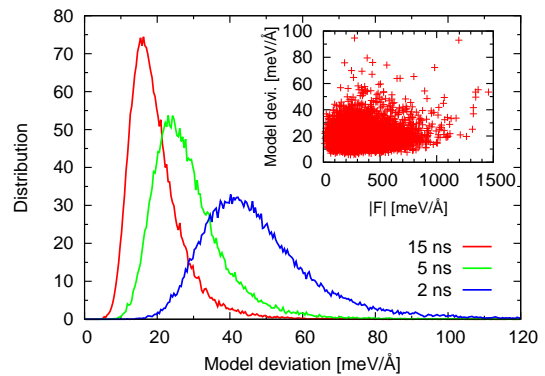}
  \caption{
    The distributions of the model deviation $\Sigma_i$ of the DeePCG model for liquid water, using training data from trajectories of total length 2~ns, 5~ns, and 15~ns.
    The 2~ns data are generated from a single trajectory;
    The 5~ns data are generated from 2 independent 2.5~ns-long trajectories;
    The 15~ns data are generated from 6 independent 2.5~ns-long trajectories.
    Inset: the correlation between the magnitude of the CG force and the model deviation, using 15~ns training data.
  }
 \label{fig:mod-dev}
\end{figure}

In this test, we prepare 1000 configurations randomly selected from the DeePMD data and pick up oxygen positions to define the CG particle configurations. 
For a CG particle $i$ in each configuration, we define the model deviation $\Sigma_i$ to be the standard deviation of the force on CG particle $i$ predicted by CG models that only differ among themselves by the initialization of the simulation procedure, i.e., 
\begin{align}
  \Sigma_i =
  \sqrt{\Big\langle
  \Vert
  \nabla_iU^{\bm w} - \big\langle\nabla_iU^{\bm w}\big\rangle_{\bm w}
  \Vert^2
  \Big\rangle_{\bm w}},
\end{align}
where the ensemble average $\langle\cdots\rangle_{\bm w}$ is taken with respect to
models obtained from the same training process, the same training data set~\footnote{The training data set is the same, but the batches used at each step of the Adam iteration are picked from the data set randomly and independently for different models.}, but different initialization of the parameters $\bm w$.
In this way, 64,000  instances of the model deviation $\Sigma_i$ are computed,
and they are used to show the consistency of the predictions of different DeePCG models quantitatively.
As shown by Fig.~\ref{fig:mod-dev}, with DeePMD data corresponding to 6 independent 2.5~ns-long trajectories, 
99.3\% of the model deviations, i.e., a large majority of them, are below 50~meV/\AA.
Moreover, the deviations do not become more significant when the magnitude of the CG force is large (inset in Fig.~\ref{fig:mod-dev}). 
Therefore, the differences of the CG forces predicted by different DeePCG models are generally consistent.
Indeed the configurational distribution functions generated by DeePCG models that differ only in the initialization are indistinguishable.
If we use shorter trajectories,
the model deviations increase, as shown in Fig.~\ref{fig:mod-dev}
for DeePMD data corresponding to 2 independent 2.5~ns-long trajectories,
and for DeePMD data corresponding to a single 2~ns-long trajectory.
This confirms that longer trajectories give better approximations of the ensemble average for $L^{\ins}(\bm w)$.

In terms of computational cost and scalability, in the current implementation, DeePCG accelerates DeePMD 7.5 times.
Since all the physical quantities in DeePCG are sums of local contributions, upon training, the DeePCG model can be directly applied to much larger systems with linear scaling of cost.
To test the reliability of DeePCG for larger systems, we perform a 1~ns-long NVT CGMD simulation on a system containing 512 water beads. 
This system is at the same temperature of the original DeePMD data, but is 8 times larger than the system used to construct the DeePCG model.
The corresponding RDF, as shown in Fig.~\ref{fig:rdf}, is only very slightly less structured than the DeePCG result with 64 water beads,
but tends to unity at large separation with a longer tail as we expect. 
This is consistent with the result in Ref. \cite{kuhne2009static}, which shows that the pair correlation function is almost converged in a 64-water fixed-cell system and
larger cells only loosen the structure very slightly.

\recheck{
Finally, like in the case of the Deep Potential and DeePMD schemes~\cite{han2017deep,zhang2018deep}, 
in our implementation discontinuities are present in the forces, due to adoption of a sharp cutoff radius,
limitation of angular information to a fixed number of atoms, and abrupt changes in the atomic lists due to sorting. 
Upon training, these discontinuities become much smaller than the thermal fluctuations and can be subsumed in the thermal noise applied by the stochastic thermostat used to sample the canonical ensemble. 
The accuracy of the canonical distributions of the CG variables reported in Figs.~\ref{fig:rdf}, \ref{fig:adf}, and \ref{fig:qs},  relative to the corresponding canonical distributions in the underlying atomistic simulation, validates our approach. 
While irrelevant for canonical sampling, the discontinuities make the CG potential a piece-wise continuous function of the CG coordinates, whereas in principle it should be a fully continuous function. 
We have recently generalized the present neural network representation in order to construct potentials that are fully continuous both in the space of the microscopic variables and in that of the CG degrees of freedom~\cite{zhang2018end}. 
We leave to future work a discussion of applications using the fully continuous (smooth) version of our approach.}

\section{Conclusion and future work}
In summary, DeePCG is a promising tool for parameterizing the CG potential and  sampling CG configurations via MD.
Due to the generality of the procedure adopted to construct the CG potential function, we expect DeePCG to be useful for a wide variety of tasks.
In the case of water, we note that one reason for the great success of the mW potential is that it allows us to accelerate ice nucleation by several orders of magnitude
because the absence of the hydrogen coordinates in the CG coordinate set eliminates the constraint imposed by the Pauling ice rules~\cite{molinero2009water}. 
It would be interesting to investigate whether the CG water model introduced in this paper could describe not only the liquid but also the crystalline ice phase, 
and whether the freezing temperature of the CG model could approximate closely that of the underlying microscopic model.
Direct ice nucleation studies would be greatly facilitated by the CG model.

Coarse grained models are often used to describe the conformations of polymers, represented for example
 by a sequence of beads and springs. 
Until  now these models are typically constructed phenomenologically by requiring that a small set of force constants match experimental and/or molecular simulation data. 
The DeePCG model presented here has the potential to completely eliminate phenomenological assumptions
 such as the restriction to harmonic spring interactions, 
by  systematically constructing a many-body potential for the beads depending on their configurations. 
We leave these studies and a more rigorous investigation of the dynamical properties of the CG models to future work.

\begin{acknowledgments}
The work of L. Zhang, J. Han, and W. E is supported 
in part by ONR grant N00014-13-1-0338, DOE grants DE-SC0008626 and DE-SC0009248, 
and NSFC grants U1430237 and 91530322. 
The work of R. Car is supported in part by DOE-SciDAC grant DE-SC0008626.
The work of H. Wang is supported by the National Science Foundation 
of China under Grants 11501039 and 91530322, 
the National Key Research and Development Program of China 
under Grants 2016YFB0201200 and 2016YFB0201203, 
and the Science Challenge Project No. JCKY2016212A502.
\end{acknowledgments}


\newpage
\appendix
\section{Definition of the local averaged Steinhardt parameters}
\label{app:a}
The bond orientational order of particle $i$ (atom or molecule) in a condensed environment is often described by a local Steinhardt parameter $q_l(i)$~\cite{steinhardt1983bond}, defined as
\begin{align}
  q_l (i)
  =
  \Big[
  \frac{4\pi}{2l+1}
  \sum_{m=-l}^{m=l} \vert q_{lm}(i)\vert^2
  \Big]^{1/2},
\end{align}
with
\begin{align}
  q_{lm}(i) =                                                                    
  \frac                                                                          
  {\sum_{j\in N_b(i)} s(r_{ij}) Y_{lm}(\hat {\vect r}_{ij})}               
  {\sum_{j\in N_b(i)} s(r_{ij})}.                                          
\end{align}
Here $N_b(i)$ denotes the set of neighbors of particle $i$, $Y_{lm}(\hat {\vect r}_{ij})$ are spherical harmonics, and $s(r_{ij})$ is a switching function defined by
\begin{align}                                                                    
  s(r) =                                                                         
  \left \{                                                                       
  \begin{aligned}                                                                
    &1, & \quad  &r < r_{min}, \\                                                 
    &\frac12 + \frac12 \cos \Big(\pi \, \frac{r - r_{min}}{r_{max} - r_{min}} \Big), & \quad &r_{min}\leq r < r_{max},  \\                                        
    &0,  & \quad  &r \geq r_{max}.                                                
  \end{aligned}                                                                  
  \right.                                                                        
\end{align}
In this work we take $r_{min} = 0.31$~nm and $r_{max} = 0.36$~nm, and adopt the modification of the local Steinhardt parameter proposed by Lechner and Dellago~\cite{lechner2008accurate},
which is more sensitive than the original bond order parameter in distinguishing different crystal structures. 
The modified Steinhardt parameter is defined by 
\begin{align}
    \bar q_l(i) = 
  \Big[
  \frac{4\pi}{2l + 1}
  \sum_{m=-l}^l \vert \bar q_{lm}(i)\vert^2
  \Big]^{\frac12},
\end{align}
with
\begin{align}
  \bar q_{lm}(i) =
  \frac
  {\sum_{j\in \tilde N_b(i)} s(r_{ij}) q_{lm}(j)}
  {\sum_{j\in \tilde N_b(i)} s(r_{ij})},   
\end{align}
where $\tilde N_b(i)$ includes $N_b(i)$ and the tagged particle $i$. 
In the full expansion of the local averaged Steinhardt parameters,
4-body terms like $Y_{lm}(\hat {\vect r}_{ik}) \cdot Y_{lm}(\hat {\vect r}_{jl})$, $i\neq j\neq k\neq l$ are found.
Therefore, the distribution of the value of the local averaged Steihardt parameters includes the effect of 4-body angular correlations.

\end{document}